\documentclass{bmcart}

\usepackage{amsthm,amsmath}
\usepackage[utf8]{inputenc} 
\usepackage{color}
\usepackage{graphicx}
\usepackage{hyperref}
\hypersetup{
    colorlinks = true,
    linkcolor = blue,
    citecolor = blue,
    urlcolor  = blue
}
\usepackage{amsmath,bm}
\usepackage{array,longtable}
\usepackage{epstopdf}

\usepackage{siunitx}
\definecolor{declared-color}{rgb}{0,0,0} 

\startlocaldefs
\usepackage{tikz}
\usetikzlibrary{calc}
\usetikzlibrary{matrix}
\usetikzlibrary{decorations.markings}
\usetikzlibrary{decorations.pathreplacing}
\usetikzlibrary{shapes}
\usetikzlibrary{shadows}
\usetikzlibrary{arrows}
\usepackage{amssymb}
\usepackage{rotating}
\usepackage{booktabs}
\usepackage{tabularx}
\usepackage{listings}
\usepackage{colortbl}
\usepackage{lscape}
\endlocaldefs

\begin{document}

\begin{frontmatter}

\begin{fmbox}
\dochead{Review}


\title{Digital Transformation in Airport Ground Operations}


\author[
   addressref={aff1},                   
   email={ivan.kovynyov@kobaltblau.com}   
]{\inits{IK}\fnm{Ivan} \snm{Kovynyov}}
\author[
   addressref={aff2},                   
   corref={aff2},                   
   email={ralf.mikut@kit.edu}   
]{\inits{RM}\fnm{Ralf} \snm{Mikut}}


\address[id=aff1]{
  \orgname{kobaltblau Management Consultants AG}, 
  \street{Bleicherweg 10},                     %
  \postcode{8002}                                
  \city{Zurich},                              
  \cny{Switzerland}                                    
}

\address[id=aff2]{
  \orgname{Institute for Automation and Applied Informatics, Karlsruhe Institute of Technology}, 
  \street{Hermann-von-Helmholtz-Platz 1},                     %
  \postcode{76344}                                
  \city{Eggenstein-Leopoldshafen},                              
  \cny{Germany}                                    
}


\begin{artnotes}
\end{artnotes}

\end{fmbox}


\begin{abstractbox}

\begin{abstract} 
How has digital transformation changed airport ground operations? Although the relevant peer-reviewed literature emphasizes the role of cost savings as a key driver behind digitalization of airport ground operations, the focus is on data-driven, customer-centric innovations.
This paper argues that ground handling agents are deploying new technologies mainly to boost process efficiency and to cut costs. Our research shows that ground handling agents are embracing current trends to craft new business models and develop new revenue streams.
In this paper, we examine the ground handling agent's value chain and identify areas that are strongly affected by digital transformation and those that are not. We discuss different business scenarios for digital technology and link them with relevant research, such as automated service data capturing, new digital services for passengers, big data, indoor navigation, and wearables in airport ground operations.
We assess the maturity level of discussed technologies using NASA technology readiness levels.
\end{abstract}




\end{abstractbox}
%

\end{frontmatter}




\section{Introduction}
Digital transformation has been described in various industries, such as
travel \cite{Fonzone16},
tourism \cite{Minghetti2010},
medical \cite{Agarwal2013},
insurance \cite{Singh2015},
consumer \cite{Maenpaa2015},
high-tech \cite{SAP2014},
energy \cite{Hagenmeyer2016, Fang2012},
public sector \cite{West2012}, and
education \cite{Kahkipuro2015}.
It appears, however, that there are a paucity of published in-depth, systematic analyses that examine airport ground operations; although it might have relevant potential for more efficient and sustainable processes (e.g., by reduced fuel consumption \cite{Singh15}) and an improve customer experience \cite{Wattanacharoensil16}.
The objective of this paper is to fill this gap by providing a resource that contains a comprehensive review of major research directions, methods and applications focused on the digital transformation of airport ground operations.

Our findings are organized as follows: First, we construct a working definition of the term digital transformation. Second, we provide an overview of an airport ground operations value chain. Next, we present a comprehensive collection of references that consider the digital transformation of airport ground operations, classifying these papers according to their respective business processes within the value chain. We discuss a real business scenario of integrated invoicing at Swissport International Ltd. We conclude this paper by highlighting key observations and offering recommendations for future research.

\section{A working definition of digital transformation}
\label{sec:digitaltrans}
Digital transformation, also referred to as digitalization \cite{GartnerITglossary}, has become a very popular term in academia and industry, yet it lacks a clear definition. In our research, we attempted to draw a clear boundary between digital transformation and simple application of information and communication technology (ICT). To this purpose, we developed a working definition and used this to select papers for our review.

New digital technologies are seen as a key driver of digital transformation, but there is no consensus about the impact of digital transformation on business. Some researchers argue that digital transformation uses technology to improve the performance of enterprises \cite{Westerman2011} and to enable major business improvements \cite{Capgemini2015}. Other papers suggest that digital transformation facilitates changes in business models, provides new revenue opportunities \cite{GartnerITglossary}, and creates new digital businesses \cite{McDonald2012}.
For this paper, we decided to use a definition that reflects the general tenor of the current debate. We defined digital transformation as \textit{the use of new digital technologies, such as cloud, mobile, big data, social media and connectivity technologies, to improve customer experience, streamline operations or create new business models}.

We did not review papers that are related to, though not strictly part of, digital transformation unless they contained at least one direct application of new digital technologies -- either to improve customer experience, streamline operations or to create new business models. Returning to the definition, let us now look at each of four main elements that we considered:

\textbf{Usage of new digital technologies}:
Current industry publications agree that major technologies accelerating digital transformation are cloud, mobile, big data, social media, and connectivity, especially smart sensors and internet of things \cite{Microsoft2015, Westerman2011}.
In our research, we relied on the following definitions of these technologies:
Cloud computing provides shared computer processing resources on demand \cite{Hassan2011, Dinh2011}.
Mobile technology involves using mobile devices to access mobile applications, data, and to communicate \cite{Microsoft2015}.
Big data are high-volume, high-velocity and high-variety data available in structured and unstructured form that require new ways of information processing for enhanced analyses and decision making \cite{Laney2001}.
Social media cover technologies that facilitate social interactions and are enabled by Internet or mobile device, such as wikis, blogs, social networks, and web conferencing \cite{Microsoft2015}.
Smart sensors are inexpensive wireless sensors of small size with an on-board microprocessor \cite{Spencer2004}.

\textbf{Improving customer experience}:
Customer service has changed and shifted towards digital self-service, such as self-service banking (funds transfer, account history, bill payments), self-service gasoline stations, self-service scanning and checkout lanes at grocery stores, electronic voting, and self-service check-in kiosks at airports \cite{Castro2010}. Self-service technology offers a broad set of benefits to consumers: it is available outside of regular business hours, saves time, protects privacy, and other benefits. Businesses invest in self-service technology to reduce costs and free up workers from routine transactions \cite{Castro2010}.
Despite these various benefits, the current literature has raised concerns about technology-based self-service: self-service simply shifts the work to the customer, reduces employment opportunities, eliminates both customer choice and human contact \cite{Reinders2008}, and is often inaccessible for elderly and disabled consumers \cite{Petrie2014}. In this review, we provided some examples of self-service technologies in airport ground operations. We have paid particular attention to concerns associated with using these technologies.

\textbf{Streamlining operations}:
Companies historically have used automation to make processes more efficient and reliable, such as in the areas of enterprise resource planning \cite{Shehab2004, Samaranayake2009}, manufacturing, research and development. Presently, process automation is often used in context of digital transformation. Industry publications report that companies drive digital transformation by automating internal processes \cite{Westerman2011, Brown2012}. Is this just another case of \textit{d\'ej\`a vu}?
Earlier papers claim that process automation leads to fundamental changes in the business \cite{Venkatraman1994}.
More recent publications argue that digital transformation has impacts beyond simply the automation of existing tasks and activities: it provides new digital services and creates new business processes. Unlike automation (modernization) in the past decades, which has focused on technology application and management, current digital transformation addresses the effects and implications of technological change \cite{Lynch2000}.
In this review, we included scientific papers that proposed how to embrace digital transformation and industry publications, showing how companies respond to digital transformation by automating internal processes.
We did not consider papers reporting simple application of ICT technology in process automation, such as automating workflows, going paperless or simply adding a digital representation of written information (digitizing information), or reducing human factors.

\textbf{New business models}: Companies apply new technologies not only to their core business, but use them to find new profit pools \cite{Brown2012, Bower1995}. In this context, digital transformation is often used as a framework with digital disruption.
Digital disruption is a technological innovation that exerts a negative impact on existing industries and creates new business opportunities \cite{Bower1995}. The literature reports several examples of how digital disruption has changed media \cite{Gilbert2012, Karimi2015}, financial \cite{Ondrus2005, Dermine2016}, and consumer \cite{Rigby2011} industries in the past decades.

\section{Value chain of airport ground operations}
\label{sec:ValueChain}
\begin{figure*}[hbtp]
\centering
\includegraphics[width=\linewidth]{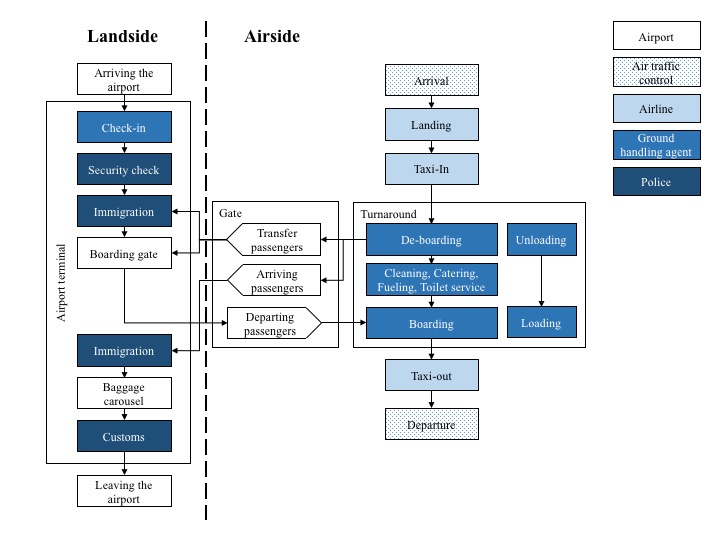}
\caption{Key airport procedures for passenger transport \cite{Kovynyov2016}}
\label{fig:airport-procedures}
\end{figure*}

Airport ground operations, also referred to as ground handling, cover those services required by an airline between landing and take-off of the aircraft, such as marshaling of aircraft, (un)loading, refueling, cleaning, catering, baggage handling, passenger handling, cargo handling, aircraft maintenance, and aviation security services.
In Figure \ref{fig:airport-procedures}, the role of the ground handling agent is shown within the scope of normal airport procedures for passenger service. The services are grouped by landside (prior to clearing security) and airside (after clearing security).
The value contribution of ground handling agents within the overall aviation value chain can be summarized as preparing the aircraft from its ground-time until the next flight.

In Figure \ref{fig:valuechain}, we present a value chain \cite{Porter1985, Porter2001} that illustrates the key processes. These processes provide service to the customer. The value chain can be divided into five core processes for airport ground handling operations.

Core business processes relate directly to the creation and delivery of ground handling services and reflect the essential functions of any ground handling agent:
\begin{itemize}
\item Passenger handling consists of the arrival (passenger de-boarding, transfer, baggage delivery, lost \& found services, arrival lounge) and departure processes (ticketing and reservation, check-in, waiting area, lounge, boarding).
\item Aircraft preparation involves parking, (un)loading, load control, fueling, pushback and towing, deicing and others.
\item Baggage handling covers, for instance, baggage drop-off, x-raying, sorting, load planning, loading and transportation.
\item Cargo handling consists of such processes as (un)loading, customs clearance, x-raying, storage and others.
\end{itemize}
Supporting processes facilitate execution of core processes:
\begin{itemize}
\item Planning \& scheduling process involves subprocesses like demand planning, shift planning, rostering (i.e. scheduling periods of duty and assignment of employees to particular shifts), daily personnel disposition and task scheduling.
\item Human resources (HR) management and training deals with employee recruitment, training demand planning, training preparation and delivery, quality assurance, payroll and dismissal.
\item Commercial processes cover negotiations of prices, tariffs and service level agreements (SLA), customer relationship management and contract entry.
\item Financial processes typically involve subprocesses such as service data capturing, invoicing, financial accounting (accounts payable, accounts receivable, balance sheet, asset accounting), financial statements, management accounting, and treasury.
\item Management processes include target setting, monitoring, reporting of key performance indicators (KPI), incentives, and leadership.
\item Procurement process focuses on buying ground support equipment (GSE) and consulting services.
\item IT process consists of solution delivery, service operations and support, and controlling.
\end{itemize}
\begin{figure*}
\footnotesize
\begin{tikzpicture}[table/.style={
  matrix of nodes,
  row sep=-\pgflinewidth,
  column sep=-\pgflinewidth,
  nodes={rectangle,text width=1.6cm,align=center},
  text depth=1.25ex,
  text height=2.5ex,
  nodes in empty cells}]

\matrix (mat) [table]
{
    ||  & ||  & ||  & ||  & ||  &   \\
    ||  & ||  & ||  & ||  & ||  &    \\
    ||  & ||  & ||  & ||  & ||  & ||  \\
    ||  & ||  & ||  & ||  & ||  & ||   \\
    ||  & ||  & ||  & ||  & ||  & ||   \\
    ||  & ||  & ||  & ||  & ||  & ||  \\
    ||  & ||  & ||  & ||  & ||  &    \\
    ||  & ||  & ||  & ||   & ||  &  \\
};

\foreach \row in {2,3,4}
  \draw[black] (mat-\row-1.north west) -- (mat-\row-6.north east);
\draw[black] (mat-5-1.north west) -- (mat-5-6.north east);

\foreach \row in {1}
  \draw[black] (mat-\row-1.north west) -- (mat-\row-5.north east);

\foreach \row in {8}
  \draw[black] (mat-\row-1.south west) -- (mat-\row-5.south east);

\foreach \col in {1,2,3,4,5}
  \draw[black] (mat-5-\col.north west) -- (mat-8-\col.south west);

\foreach \col in {1}
  \draw[black] (mat-1-\col.north west) -- (mat-5-\col.south west);

\node at (mat-1-3) {Planning \& Scheduling};
\node at (mat-2-3) {HR \& Training};
\node at (mat-3-3) {Commercial Process};
\node at (mat-4-3) {Financial, Management, Procurement, IT Processes};
\node at ([yshift=-10pt]mat-6-1) {\parbox[t]{2cm}{\centering Passenger Arrival}};
\node at ([yshift=-10pt]mat-6-2) {\parbox[t]{2cm}{\centering Aircraft Preparation}};
\node at ([yshift=-10pt]mat-6-3) {\parbox[t]{2cm}{\centering Baggage Handling}};
\node at ([yshift=-10pt]mat-6-4) {\parbox[t]{2cm}{\centering Cargo Handling}};
\node at ([yshift=-10pt]mat-6-5) {\parbox[t]{2cm}{\centering Passenger Departure}};
\node[rotate = 90] at ([xshift=-52pt]mat-3-1.north) {Supporting Processes};
\node at ([yshift=-19pt,xshift=-0.5cm]mat-8-3.south) {Core Processes};

\fill[white] (mat-1-5.north east) -- (mat-5-6.north east) -- (mat-1-6.north east) -- cycle;
\fill[white] (mat-8-5.north east) -- (mat-5-6.north east) -- (mat-8-6.north east) -- cycle;

\shade[top color=white,bottom color=white,middle color=white,draw=black]
  (mat-1-5.north) -- (mat-5-6.north) -- (mat-8-5.south) --
  (mat-8-5.south east) -- (mat-5-6.north east) -- (mat-8-5.south east) --
  (mat-5-6.north east) -- (mat-1-5.north east) -- cycle;

\begin{scope}[decoration={markings,mark=at position .5 with \node[transform shape] {Margin};}]
\path[postaction={decorate}]
  ( $ (mat-1-5.north)!0.5!(mat-1-5.north east) $ ) -- ( $ (mat-5-6.north)!0.5!(mat-5-6.north east) $ );
\path[postaction={decorate}]
  ( $ (mat-5-6.north)!0.5!(mat-5-6.north east) $ ) -- ( $ (mat-8-5.south)!0.5!(mat-8-5.south east) $ );
\end{scope}

\draw[decorate,decoration={brace,mirror,raise=6pt}]
  (mat-1-1.north west) -- (mat-5-1.north west);
\draw[decorate,decoration={brace,mirror,raise=6pt}]
  (mat-8-1.south west) -- (mat-8-5.south);
\end{tikzpicture}

\caption{Value chain of airport ground operations}
\label{fig:valuechain}
\end{figure*}

\section{Applications}
\label{sec:Applications}
The papers for this section were selected based on our working definition of digital transformation, which we described previously. We categorized each paper according to the process of the ground handler's value chain involved (see Table \ref{tab:summary-core} for core processes and Table \ref{tab:summary-support} for supporting processes).

For the sake of simplicity, passenger arrival and passenger departure are referred to as passenger handling.

We carried out the search for papers using a wide range of electronic libraries across the world, electronic journal collections, general web searches, communication with authors and interviews with industry experts (airlines, airports, ground handling companies).

Based on our interviews, we identified further examples of digital transformation in airport ground operations; however, we were unable to locate papers relevant to these scenarios. We have included these examples in Table \ref{tab:summary-core} and Table \ref{tab:summary-support}, and introduced selected scenarios in detail in this section.
\begin{table*}[htp]
\caption{Key areas of digital transformation of airport ground operations (core processes)}

 \footnotesize

\begin{center}
\renewcommand{\arraystretch}{0.9}

 \begin{tabular}{ p{2cm}p{3cm}cccp{5cm} }

\hline
Business process 		& Business scenario 						& CUS 	& OPS 	& BIZ 	& Authors \\
\hline



Passenger handling 		& Indoor navigation 						& * 		&  		&  * 		&
Darvishy et al. \cite{Darvishy2008},
Fallah et al. \cite{Fallah2013},
Hat el al. \cite{Han2014},
Odijk and Kleijer \cite{Odijk2008},
Radaha et al. \cite{Radaha2015}\\

					& Digital processing of irregularity vouchers 	& * 		& * 		& 		&
McCollough et al. \cite{McCollough2000}\\

					& Self-service check-in kiosks 				& * 		& * 		&		&
Abdelaziz et al. \cite{Abdelaziz2010},
Castillo-Manzano et al. \cite{Castillo-Manzano2013},
Chang and Yang \cite{Chang2008},
Howes \cite{Howes2006},
Ku and Chen \cite{Ku2013},
Liljander et al.\cite{Liljander2006},
Wittmer \cite{Wittmer2011a}\\

					& Self-boarding							& * 		& * 		&		& Jaffer and Timbrell \cite{Jaffer2014} \\
					& Smart wheelchairs						& *		& * 		&		&
Berkvens et al.\cite{Berkvens2012},
Morales et al. \cite{Morales2014},					
Romero et al. \cite{Romero2009} \\
					& Smart wearables						& * 		& * 		&		& n.a. \\
					& Biometric services						& * 		& * 		&		&
Oostveen et al. \cite{Oostveen2014},
Palmer and Hurrey \cite{Palmer2012},
Sumner \cite{Sumner2007},
Sasse \cite{Sasse2002},
Scherer and Ceschi \cite{Scherer2000}
\\ 					
\hline
Baggage handling 		& RFID baggage tags 					& * 		& * 		&		&
Berrada and Salih-alj \cite{Berrada2015},
Bite \cite{Bite2010},
Datta et al. \cite{Datta2016},
DeVries \cite{DeVries2008},
Mishra and Mishra \cite{Mishra2010},
Zhang et al. \cite{Zhang2008}\\
					& Automated baggage drop-off				& *		& *		& 		&
Jaffer and Timbrell \cite{Jaffer2014},
Wittmer \cite{Wittmer2011a}\\
					& Digital bag tags						& *		& * 		& 		& n.a. \\
					& Self-tagging							& *		& * 		& 		& n.a. \\		
					& Lost luggage kiosks					& *		& *		&		& n.a. \\	
					& Real-time luggage tracking				& *		& *		&		&
Sennou et al.\cite{Sennou2013} \\	
					& Advanced analytics					& *		& *		&		& n.a. \\						
							
\hline


Lounge services 		& Lounge access gates 					& *		& *		& 		& n.a. \\
					& New lounge ticket types					& *		&		& *		& n.a. \\
\hline
\end{tabular}
\end{center}
* Business scenario addresses:
CUS = improving customer experience,
OPS = streamlining operations,
BIZ = creating new business models.
\label{tab:summary-core}
\end{table*}
%
%
\begin{table*}[htp]
\caption{Key areas of digital transformation of airport ground operations (supporting processes)}

 \footnotesize

\begin{center}
\renewcommand{\arraystretch}{0.9}

 \begin{tabular}{ p{2cm}p{3cm}cccp{5cm} }

\hline
\rowcolor[gray]{.9}
Business process 		& Business scenario 						& CUS 	& OPS 	& BIZ 	& Authors \\
\hline

Planning \& Scheduling 	& Automated centralized planning			& 		& *		&		& Herbers \cite{Herbers2005},
Herbers and Kutschka \cite{Herbers2016},
Ernst et al. \cite{Ernst2004},
Ip et al. \cite{Ip2010},
Ernst et al. \cite{Ernst2004a},
Dowling et al. \cite{Dowling1997},
Mason and Ryan \cite{Mason1998},
Stolletz and Zamorano \cite{Stolletz2014},
Brusco et al. \cite{Brusco1995},
Stolletz \cite{Stolletz2010},
Chu \cite{Chu2007},
Dorndorf \cite{Dorndorf2006},
Keller and Kruse \cite{Keller2002}
\\
					& De-peaking							& 		& *		&		& Kisseleff and Luethi \cite{Kisseleff2008}, Luethi and Nash \cite{Luethi2009} \\
					& Shift trading 							& 		& *		& *		& n.a. \\

\hline
GSE management 		& Automated GSE scheduling and routing				&		& *		& 		& Padron et al. \cite{Padron2016}, Norin et al.  \cite{Norin2012}, Kuhn and Loth \cite{Kuhn2009} \\
\hline

HR \& Training 			& Digital employee profiles 				& 		& * 		& 		& n.a. \\
					& Web-based training 					& 		& * 		& 		& n.a. \\
					& Integrated employee lifecycle management 	& 		& * 		& 		& n.a. \\
					& Mobile apps and social-media for corporate communications & & * & & n.a. \\
\hline
Financial process 		& Integrated invoicing 					& 		& * 		& 		& Kovynyov et al. \cite{Kovynyov2010, Kovynyov2016} \\
\hline
Management process 	& Integrated KPI reporting					& 		& * 		& 		& Schmidberger et al. \cite{Schmidberger2009} \\
\hline
IT process 			& Digital workplace and virtualization 		& 		& *		& 		& n.a. \\
\hline
\end{tabular}
\end{center}
* Business scenario addresses:
CUS = improving customer experience,
OPS = streamlining operations,
BIZ = creating new business models.
\label{tab:summary-support}
\end{table*}

\subsection{Passenger Handling}
\begin{figure}[hbt]
\centering
\includegraphics[width=0.9\columnwidth]{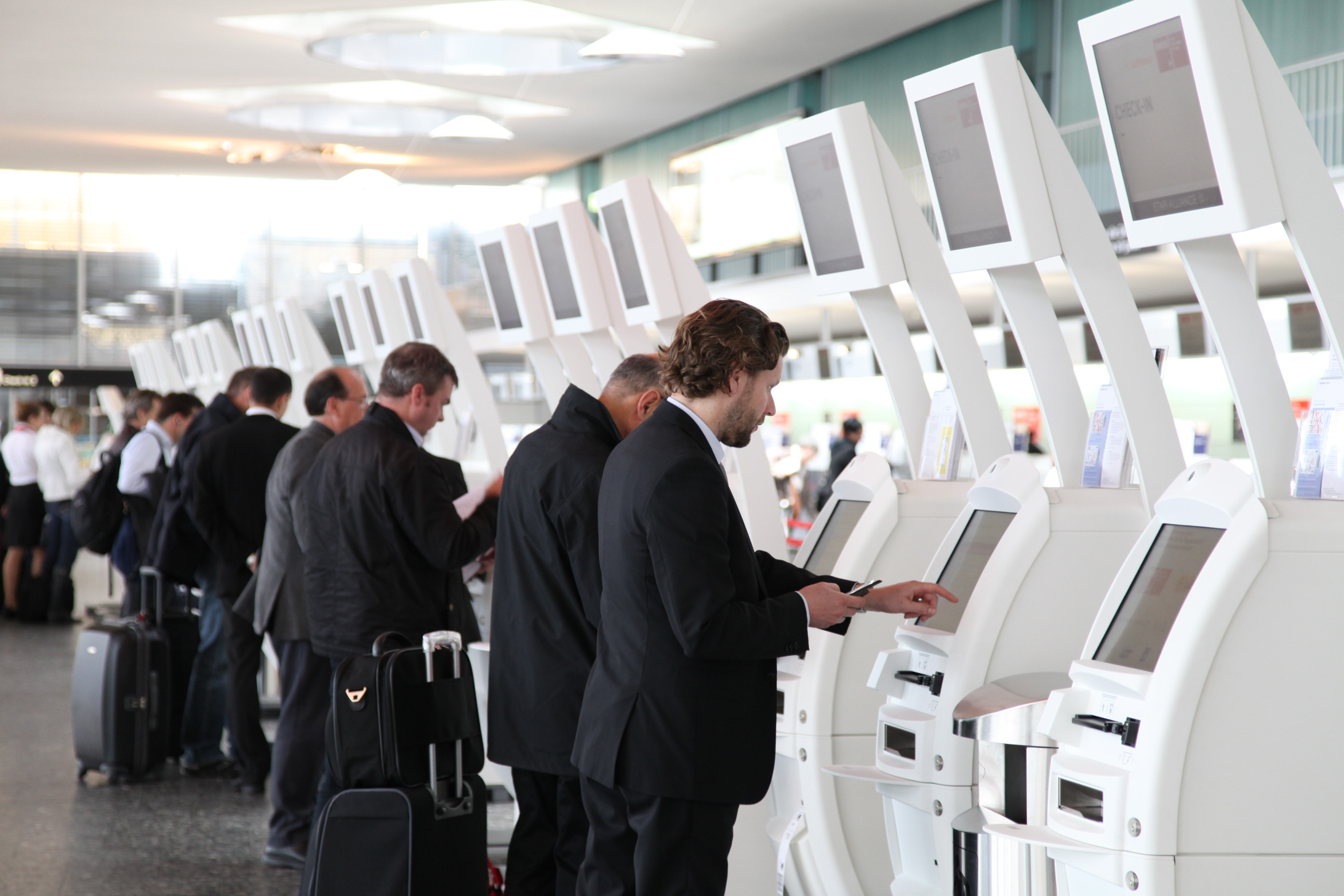}
\caption{Self-service check-in at Zurich airport (Photo courtesy Swissport)}
\label{fig:cuss}
\end{figure}

\textbf{Self-service check-in kiosks} are computer terminals for passenger check-in which remove the need for ground staff (Fig. \ref{fig:cuss}). Previously, airlines had only dedicated kiosks for their own passengers; today,  kiosks share check-in applications for multiple airlines \cite{Howes2006}. Shared kiosks bring several benefits:  airports can better utilize the limited space in airport terminals, airlines eliminate capital spending, since they buy these services directly from a ground handling agent, ground handling agents can cut operating costs by engaging fewer ground staff for passenger check-in, and passengers can save time as they can use any kiosk instead of having to search for dedicated kiosks. Ground handling agents install self-service check-in facilities and promote them by engaging floor-walkers to assist passengers in the land-side of check-in areas \cite{Wittmer2011a}.

\textbf{Self-boarding}:
Quick boarding gates allow passengers to self-scan the boarding pass at the gate. After the boarding pass has been verified, the gates are released and the passenger can proceed to the aircraft. In this arrangement,  ground staff are not involved into the passenger boarding and can focus on supervisory tasks or special cases. This helps reduce the need for ground staff, and ground handling agents are able to further reduce operating costs.
Current literature argues that self-boarding has a positive impact on customer satisfaction, because it significantly reduces processing times at the gate \cite{Jaffer2014}. Quick boarding gates also record various operational data which can offer more insight into customer behavior or efficiency of the boarding process, such as processing times by flight and gate, passenger groups using boarding gates, and passenger distribution over the time.

\textbf{Indoor navigation:}
Many recently published papers study indoor navigation in airport terminals \cite{Han2014}, indoor navigation for passengers with reduced mobility \cite{Darvishy2008}, for transit passengers \cite{Radaha2015}, and location-based services at airports \cite{Odijk2008}. An overview of the development of this technology can be traced through earlier review papers \cite{Fallah2013}.
Currently, ground handling agents provide special assistance services for elderly passengers, those with reduced mobility, and unaccompanied minors. The impact of indoor navigation on this aspect of the business has not yet been described in the literature. Indoor navigation technologies have the potential to disrupt the market for airport assistance services, if passengers adopt this technology and instead use a mobile device to navigate, rather than booking special assistance services.

\textbf{Smart wheelchairs}:
Recent papers report some applications of autonomous, self-driving GSE vehicles \cite{Morris2015}. Further applications of autonomous driving technologies are reported in automobile, truck, public transportation, industrial and military services \cite{Bishop2000}.
So far, we have not found any papers on self-driving vehicles within airport terminals. However, we believe that providing self-driving cars within the airport terminal can open up new possibilities in the special assistance of passengers with reduced mobility. Currently, ground staff drive passengers using golf-carts or in wheelchairs through the airport terminal. We believe that ground handling agents can reduce the amount of ground staff employed by using self-driving, electrically powered wheelchairs or golf-carts with manual and visual control systems \cite{Romero2009}.

\textbf{Digital processing of irregularity vouchers:}
Irregularity vouchers are free cash vouchers for meals, hotel accommodation, airport transfers, and bag replacement provided to passengers by an airline (or ground handling agent on behalf of the airline) in case of a flight delay. The vouchers are usually personalized. The vouchers are redeemed from the airline and are for either pre-arranged services or a maximum indicated value. Usually, airlines have a preexisting agreement with specific hotels regarding accommodation and rates.
For meals, vouchers are in many cases presented to multiple vendors and usually include a maximum redemption value. Some merchants who accept such vouchers in bulk, redeem them from the airline using electronic data processing. Businesses that do not receive many vouchers or do not use an electronic system usually physically mail the vouchers to the airline or ground handling agent in order to receive payment. We believe that digital processing of irregularity vouchers has significantly reduced processing costs for service providers, airlines and ground handling agents.
Irregularity vouchers have been extensively studied in the literature relating to customer satisfaction after service failure and recovery \cite{McCollough2000}. However, so far, we have not found any publications describing digital processing of irregularity vouchers.

\textbf{Smart wearables}:
Smart watches, miniature wrist-mounted computers with a time-keeping functionality and an array of sensors \cite{Rawassizadeh2014}, are widely used by passengers at airports. With a smart watch, passengers can get alerts on gate changes or flight delays, scan a boarding pass at security late or at the gate. As the smart watch is permanently worn on the wrist, ground handling agents can use this channel to broadcast real-time information to passengers outside of traditional visual display monitors, voice announcements and send messages to other mobile devices.
We have some found some evidence of pilot projects with smart watches. For example, a major U.S. airline announced a trial of smart glasses and smart watches, using the products to greet passengers by name, provide real-time travel information and start the check-in process before the passenger even reached the front door of the terminal. Nonetheless, we have not found any papers that describe the usage of smart watches in the passenger handling. We propose the following for future research directions: indoor navigation at airports using smart watches for visually impaired passengers, sharing health data for cross-border disease control, and emergency notifications on smart watches.

\textbf{Biometric services}:
Biometrics are automated person identification using physiological characteristics (face, fingerprints, hand geometry, handwriting, iris, retinal, vein, voice) \cite{Heracleous2006}. Current research reports several applications of biometric technology at airports, such as
airport security,
biometric travel documents,
airport access control \cite{Sumner2007},
biometrics in security,
biometrics in baggage claim \cite{Scherer2000},
airport immigration systems \cite{Sasse2002, Oostveen2014, Palmer2012},
and seamless travel \cite{Heracleous2006}.
\subsection{Baggage Handling}

Recent innovations in the area of baggage handling are mainly driven by the following trends: airlines boost productivity by reducing ground times (timeframe between landing and take-off), seat capacity of new aircraft increases, and timely delivery of baggage became part of the service level agreements between ground handling agents and airlines. Clearly, these trends introduce new requirements to baggage handling systems and related processes in terms of correctness and speed of baggage processing. Consider now some examples how did ground handling agents respond.

\begin{figure}[hbt]
\centering
\includegraphics[width=0.5\columnwidth]{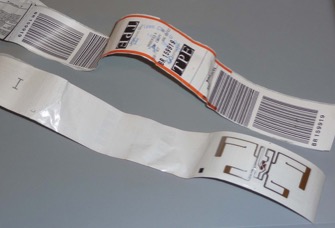}
\caption{RFID tag incorporated into a bag tag (Photo Vanderlande)}
\label{fig:rfid-bag-tag}
\end{figure}

\textbf{RFID baggage tags:}
the radio frequency identification (RFID) technology enables identification from a distance and does not require a line of sight, unlike the bar-code technology \cite{Want2006}. The benefits of RFID baggage tags have been extensively discussed in the current literature:
RFID tags can improve baggage tracing \cite{Zhang2008}, baggage routing during air transit \cite{Datta2016}, and reduce the amount of misrouted luggage \cite{DeVries2008}. Furthermore, RFID tags can incorporate additional data such as manufacturer, product type, and even measure environmental factors such as temperature \cite{Hunt2006}. RFID systems can identify many different tags located in the same general area without human assistance. Embedded in barcode labels, RFID tags could eliminate the need for manual inspections and routing by ground handling agents.
At the moment, RFID tags are inserted into paper and then attached as paper labels to the baggage (Fig. \ref{fig:rfid-bag-tag}).
%
%

\textbf{Self-tagging} is one of latest ideas for bag tagging. Passengers tag their own bags, print luggage tags at home and track their bags on smartphones. In this context, digital bag tags are gaining more importance. \textbf{Digital bag tags} are the digital alternative to conventional paper-based baggage tags. Bags receive a permanent bag tag that displays a digital bar-code. Airlines or ground handling agents are able to change this bar-code remotely, if the flight plan has changed or a passenger has been re-toured.  Combined with a tracking device, that is stored inside the bag, passengers are able to track the luggage on smartphone in real-time.

\textbf{Automated baggage drop-off:} Combined with home-printed or electronic bag tags, passengers can use fully automated machines to receive passenger-tagged bags without any interaction with ground staff or airline employees. Several airports have already installed automated bag-drop machines \cite{Jaffer2014}.

\begin{figure}[hbt]
\centering
\includegraphics[width=0.9\columnwidth]{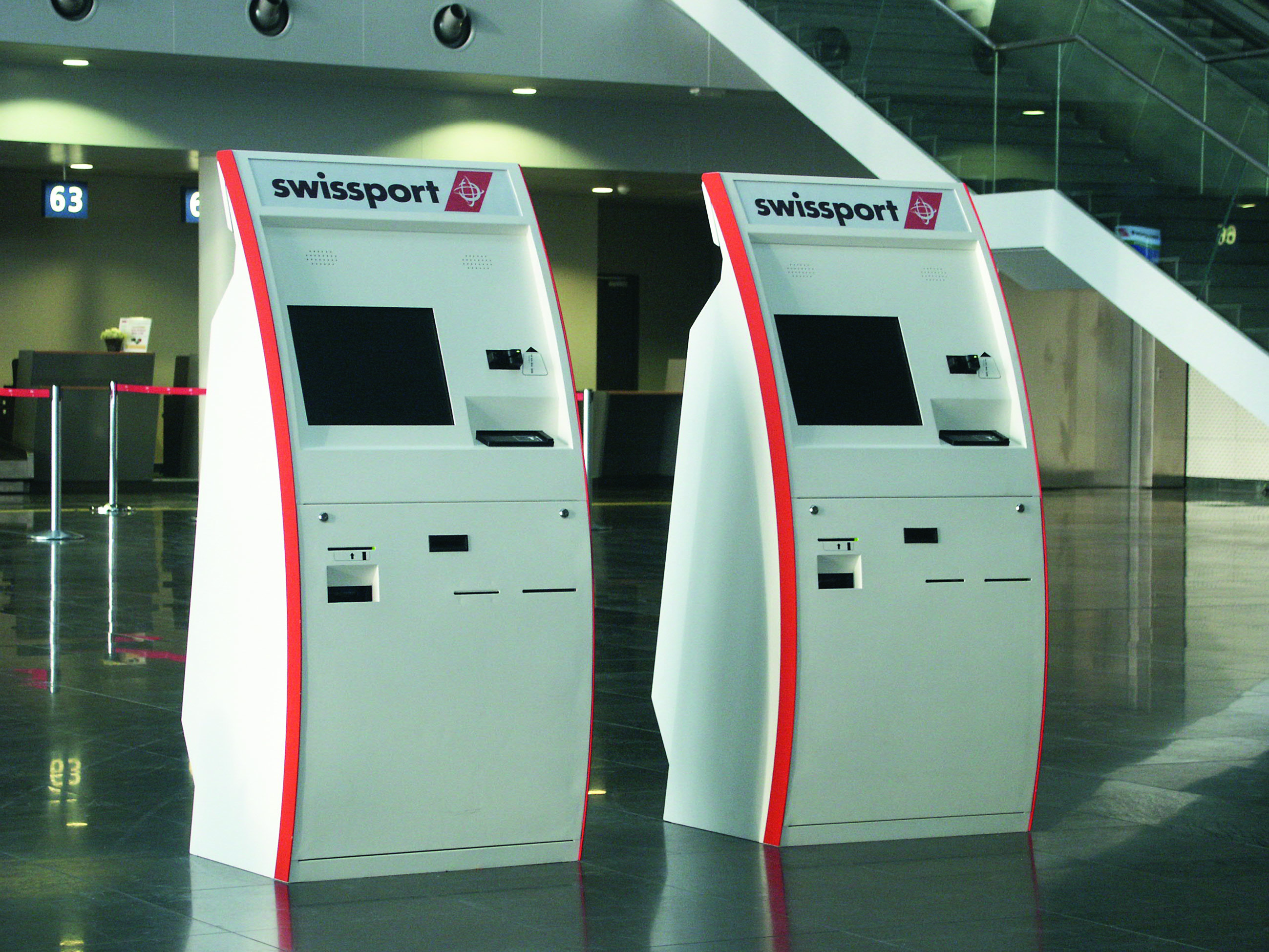}
\caption{Lost luggage kiosks at Geneva Airport (Photo courtesy Swissport International Ltd.)}
\label{fig:lost-luggage-kiosk}
\end{figure}

\textbf{Lost luggage kiosks} are self-service computer terminals for reporting lost baggage. They are connected to the global database for lost luggage and help passengers to report delayed or missing bags upon arrival (see Figure \ref{fig:lost-luggage-kiosk}). In order to report lost luggage, a passenger scans the boarding pass, describes the missing item and enters contact details for the delivery, for when the luggage has been found. Passengers can obtain the latest information to the report by accessing a website and entering the number of their report.

\textbf{Advanced analytics in baggage handling}:
Performance-related contracts between ground handling agents and airlines require reliable data on speed and efficiency of baggage handling systems, such as in-time delivery of baggage (first bag, last bag), processing times, and correctness of baggage processing. In addition, ground handling agents collect and maintain data pockets on operational baggage flow, develop simulations and models to understand past behavior and performance of baggage handling systems.

\subsection{Lounge Services}
\textbf{Lounge access gates:}  Most passengers use status-based access or class-of-travel access rather than membership. Status passengers can swipe the card in order to access a lounge. Passengers with business class or first class tickets are required to show the boarding pass to enter the lounge.
Some cardmembers may receive access privileges to the lounges. In order to access the lounge, cardmembers must present their card. Furthermore, some passengers are allowed to bring guests. Ground handling agents need to know such regulations for several airlines and lounge clubs. In addition, lounge personnel need to be able to apply these rules immediately when a particular passenger is attempting to enter a lounge. To this purpose, ground handling agents have installed automated access gates at lounge entrances. The access gate is connected to a back-end application with the rule sheet. As soon as a passenger has swiped the card or scanned the boarding pass, the application applies the rule sheet and decides if the passenger is able to access the lounge.

\textbf{New lounge ticket types}: Until recently, ground handling agents have been selling single entries to airlines or directly to passengers. Today, ground handling agents have begun introducing new types of products, such as family tickets and lounge access vouchers offered by tour operators. Family tickets can be booked in advance and offer group access to the lounge. Family tickets are sold for a fixed price irrespective of the status or booking class of the passengers. Furthermore, ground handling agents have begun forming cooperations with charter airlines and tour operators, in order to include lounge access into offered travel packages. New types of products introduce new and possible greater obligations on ground handling agent's IT systems: lounge access gates need to be coordinated with family tickets and tour operator tickets, invoicing systems need to able to price and bill the tickets appropriately, and data analytics facilities need to be able to analyze customer behavior and provide insights into the new product types.

\subsection{Staff Planning and Scheduling}
Since labor costs in airport ground operations account for 60 to 80 percent of the total cost, automation of staff planning and scheduling is a top priority for every ground handling agent \cite{Herbers2014a}. The development of research on rostering and task scheduling of airport ground staff can be traced through earlier review papers \cite{Herbers2005, Herbers2016, Ernst2004, Ip2010, Ernst2004a}.

\textbf{Automated centralized planning} is the most common approach reported in the literature. This approach is based on prior demand modeling and stepwise reduction of the planning horizon.
Usually, the planning process begins six months in advance, using the flight schedule forecast for the next season (winter or summer).
The aggregate flight schedule forecast is achieved via the production management system, and is likely to be just a distribution of flights by month, airline and destination. The aggregate flight schedule forecast is used for holiday planning and strategic workforce planning. Two months in advance, the aggregate forecast is transformed into a detailed flight schedule forecast. Based on the detailed forecast, the shift plan is compiled. The shift plan is a result of balancing demand quantified in the flight schedule forecast against the available workforce. The shift plan includes the distribution of different shifts per day throughout the month and associated skills required for performing these shifts. The shift plan is usually performed using a dedicated planning application. Next, the shift plan is transformed into the rostering plan. The rostering plan assigns employees to particular shifts and is prepared at least two weeks before its application \cite{Dowling1997, Mason1998, Stolletz2014, Brusco1995, Stolletz2010, Chu2007}. The rostering plan is usually completed using  the time and attendance system.
Hours prior, the rostering plan is incorporated with the task schedule\cite{Dorndorf2006}. The task schedule assigns particular tasks to the shifts. Usually, the task schedule is created using a dedicated real-time application with broadcasting capacity to the employees' hand-helds.

\textbf{De-peaking} is a strategy whereby air traffic between peak and off-peak times is scheduled so as to distribute traffic more evenly throughout the day. Conventional planning approaches do not include de-peaking by default. Ground handling agents operating at hub airports incorporated additional de-peaking strategies into their planning procedures \cite{Kisseleff2008, Luethi2009}.

\textbf{Shift trading} is an on-line service to exchange work shifts.
As soon as a new shift plan is announced, employees can request shift trades.
Shift trading brings many benefits to ground handling agents: it helps eliminate planning conflicts along the way, it provides a collaborative environment to employees and gives the freedom of choice (both are highly appreciated by employees, which increases employee satisfaction), and employee self-organization is more cost effective for the company.
We observed two organizational models for shift trading -- either it is incorporated into the time and attendance recording system, or results as a product of from employee self-organization, via a satellite website or closed groups in social networks. Employee-led shift trading has the potential to create new business models.

\subsection{GSE Management}
Ground support equipment (GSE) is usually found on the ramp (i.e. servicing area of airport, also referred to as the apron). GSE equipment includes, for instance, refuellers, container loaders, belt loaders, transporters, water trucks, ground power units, air starter units, lavatory service vehicles, catering vehicles, passenger boarding stairs, passenger buses, pushback tractors, de-icing vehicles, container dollies, cargo pallets, and others.

\textbf{Automated GSE scheduling} is one of the most common applications in this area. It minimizes the total number of GSE vehicles required to handle flights and, consequently, reduces ground handling agent's capital expenditures. Recent literature describes advanced scheduling methods for GSE vehicles \cite{Padron2016} and deicing trucks \cite{Norin2012}. Advanced scheduling and routing algorithms minimize the overall apron traffic and reduce ground handling agent's fuel costs \cite{Kuhn2009}. Ground handling agents use GPS localization and tracking of GSE vehicles to collect data on effective use, and use them for scheduling and forecasting.

\subsection{HR \& Training}
The data on digital transformation in the human resources (HR) \& training processes in airport ground operations are scarce. Based on our interviews with industry experts, we were able to identify some business scenarios that illustrate how digital transformation has changed the HR \& training processes:

\textbf{Digital employee profiles}: Ground handling agents have a large number of employees and need to access employee information quickly. Digital employee profiles create a centralized overview of employees by integrating multiple sources of employee-related information:
personal (age, gender, address, civil status),
financial (compensation, including rewards and benefits),
qualifications (certifications, trainings, skills, performance reviews), and
workforce data (productivity, absences, holidays, overtime balances).

\textbf{Web-based trainings}: Ground handling agents started using web-based training instead of classroom training. Web-based training involves using browser-based learning programs available on the corporate intranet. Such trainings can be accessed as desktop applications or from a mobile device (e.g. tablet or smartphone). Web-based training includes single and recurrent standard training programs (e.g. code of conduct, health and safety),
management training (e.g. anti-corruption and fair competition guidelines), and technical training for operational staff (e.g. dangerous goods, security regulations, customer service). Most web-based training programs are oriented to knowledge transfer (e.g. new regulations, rules, procedures) and do not require interaction with trainers. Web-based training frees up in-house trainers from routine tasks and reduces training costs. Furthermore, web-based training can be easily translated into other languages or enhanced using local information.

\textbf{Mobile apps and social media for corporate communications}:
Ground handling agents need to be able to distribute the latest news, reports and corporate announcements to a large number of employees in a timely fashion. Operational staff, in particular, need to be promptly informed about security deficits, bomb threats, aircraft damages, irregularities in airport operations, and other considerations. Consequently, many of ground handling agents have installed visual displays in lunch rooms and offices, developed corporate mobile apps or started engaging with employees via social media networks.

\textbf{Integrated employee lifecycle management} considers all steps an employee follows during their time within a ground handling company. This includes recruitment, on-boarding, goal setting and performance reviews, personal development, talent management, succession planning, and departure (retirement, dismissal, leave). Ground handling agents manage large numbers of employees and must be cost-sensitive. HR organizations need to be highly efficient in executing their tasks. For example, they might improve cost-effectiveness of advertising by tapping in to social networks to engage with potential candidates, publishing job openings via internal job markets, or ensure job application submissions are quick and easy to review, responding to candidates promptly, relying on cost-effective selections methods.
\begin{table}[hbt]
\caption{Key performance management ratios}

 \footnotesize

\begin{center}
\renewcommand{\arraystretch}{0.9}

 \begin{tabularx}{\columnwidth}{p{.3\columnwidth}p{.6\columnwidth}}

\hline
Operational ratios		& Total hours worked per departing flight \\
					& Total minutes worked per departing passenger  \\
\hline

Financial ratios			& Direct cost per flight departure \\
					& Labor cost per flight departure \\
					& Total revenues per flight departure \\
\hline

HR ratios				& Overhead ratio (overhead : total FTEs*)  \\
					& Absence ratio (absence hours : total work hours) \\
					& Staff turnover rate (entries \& leaves : average number of employees) \\
\hline

Safety \& quality ratios	& Employee injuries per 100 flights \\
					& On-time performance (delay minutes per 100 flights) \\
					& Aircraft damage ratio (severe aircraft damages per 100 flights)  \\
\hline

\end{tabularx}
\end{center}
* FTE = full time equivalent

Source: our analysis and Schmidberger et al. \cite{Schmidberger2009}
\label{tab:key-ratios}
\end{table}

\subsection{Management Process}
\textbf{Integrated KPI reporting}: Table \ref{tab:key-ratios} shows a summary of ratios used by ground handling agents to measure and track corporate performance. These ratios are designed to address key areas of concern, which include operations, finance, human resources, safety and quality.
The calculation of these ratios requires an intelligent combination of information from multiple sources, such as the latest worked hours from the time and attendance system (for operational rations), as well numbers of handled flights and passengers maintained in the production management system. Similarly, financial ratios require data from the payroll system, the enterprise resource planning (ERP) system and the production management system. All ratios should be calculated automatically and provided to the management at least on a daily basis. Consequently, ground handling agents invest in consolidation, standardization and active management of data pockets providing this kind of information. Advanced analytics and big data technologies create the possibility to develop further insight into the business. In our research, we were not able to find any evidence that ground handling agents are using these technologies in the context of such business scenarios.

\subsection{Integrated Invoicing: a Case Study}
\label{sec:UseCase}

We provide a real business scenario showing how digital transformation has changed financial and commercial processes in airport ground operations. The ground handling agent Swissport International Ltd. undertook a program to radically automate and standardize the invoicing process and codify customer contracts in the invoicing system  \cite{Kovynyov2010, Kovynyov2016}. The insights presented here consider the results of this program.

The program executed by Swissport is one example of digital transformation for the following reasons:
\begin{itemize}
\item new digital technologies (e.g. smart sensors, embedded devices, mobile apps) were used to establish a  connection between data capturing facilities and the invoicing system;
\item big data technologies and advanced analytics were employed to ensure data quality;
\item the financial department and technology staff collaborated to digitize customer contracts and enter all customer-related prices, tariffs and incentives into the invoicing system. These measures drove changes in business processes and systems, enabling automated processing of invoices;
\item process automation halved manual workloads and significantly reduced processing times during billing cycles;
\item new technology eased data sharing with business partners and facilitated e-billing (i.e. digital transmission of invoices to customers);
\item customers saw substantial improvements through new invoice formats and greater consistency in the way the services were presented on the invoice, leading to improved customer satisfaction and retention.
\end{itemize}
\begin{landscape}
\begin{figure}[hbtp]
\footnotesize

\begin{tikzpicture}[
thick,
>=stealth,
database/.style={
      cylinder,
     cylinder uses custom fill,
     cylinder body fill=gray!20,
     cylinder end fill=gray!50,
      shape border rotate=90,
      minimum width=1.6cm,
      minimum height=1cm,
      aspect=0.25,
      draw=gray},
widebox/.style={
      rectangle,
      minimum width=1.6cm,
      minimum height=.7cm,
      draw=gray,
      fill=gray!25},
tapebox/.style={
      double copy shadow,fill=gray!20,draw=gray,
      tape,
      minimum width=1.6cm,
      minimum height=1cm,
      draw
    }
]

\node[rectangle, draw, minimum width=3cm, minimum height=.6cm] at (0,1.3) {\textbf{Data capturing}};
\node[rectangle, draw, minimum width=4cm, minimum height=.6cm] at (3.5,1.3) {\textbf{Data consolidation}};
\node[rectangle, draw, minimum width=9cm, minimum height=.6cm] at (10,1.3) {\textbf{Issue of invoice}};

\node[database] (n0) at (0,0) {AODB};
\node[align=center] (n1) at (0,-1.5) {\includegraphics[scale=.026]{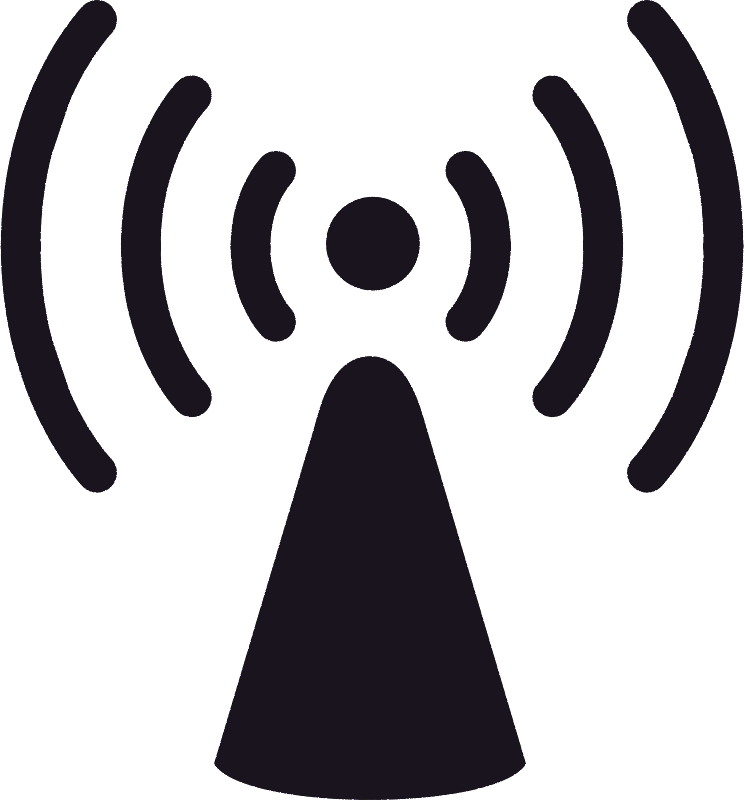} \\ Sensors};
\node[widebox] (n2) at (0,-2.7) {Mobile};
\node[tapebox] (n3) at (0,-4) {Scans};
\node[widebox] (n4) at (0,-5.3) {Manual};
\node[cloud, draw=gray, aspect=2, fill=gray!25] (n5) at (0,-6.6) {Cloud};

\node[cylinder, cylinder uses custom fill, cylinder end fill=gray!50,cylinder body fill=gray!20, minimum width=0.8cm, draw=gray] (n6) at (2.5,0) {Flights};
\node[rectangle, draw=gray, fill=gray!20, minimum width=5cm, minimum height=0.8cm, rotate=90] (n7) at (2.5,-4) {Services};
\node[database, minimum height=5cm, align=center] (n8) at (4.5,-4) {Data \\ mart};
\node[tapebox, align=center] (n18) at (4.5,-7.5) {Analyses, \\ Stats};
\node[rectangle, draw, anchor=north, minimum height=7cm, minimum width=8.5cm, label=above: ERP system] (n9) at (10,0.4) {};

\node[double copy shadow, fill=gray!20,draw=gray, rectangle, align=center, minimum height=5cm] (n10) at (6.8,-4) {Import \\tables};
\node[double copy shadow, fill=gray!20,draw=gray, rectangle, align=center, minimum height=1.5cm] (n11) at (11,-0.8) {Customer \\ contracts};
\node[rectangle, draw=gray, align=left, fill=gray!20] (n12) at (11,-4) {\textbf{Sales order:} \\ customer \\ material \\quantity \\ price};
\node[double copy shadow, fill=gray!20,draw=gray, rectangle, align=center, minimum height=1.5cm] (n13) at (8.8,-0.8) {Mapping \\ tables};
\node[circle, minimum size=1.5mm,inner sep=0pt,outer sep=0pt, fill=black, draw] (n15) at (8.8,-4) {};
\node[rectangle, draw=gray, fill=gray!20, align=center] (n16) at (13.3,-2.5) {Billing \\ doc};
\node[tape, draw=gray, fill=gray!20, minimum height=1cm] (n17) at (13.3,-4) {Invoice};
\node[align=center] (n19) at (13.3, -1) {General \\ ledger};


\path [->, thick]
(n0) edge node [above left] {} (n6)
(n1.base east) edge node [above left] {} (n7)
(n2.east) edge node [above left] {} (n7)
(n3.east) edge node [above left] {} (n7)
(n4.east) edge node [above left] {} (n7)
(n5.puff 9) edge node [above left] {} (n7)
(n7) edge node [above left] {} (n8)
(n8) edge node [above left] {} (n10)
(n13) edge node [fill=white, align=center] {customer, \\material} (n15)
(n11) edge node [fill=white] {price} (n12)
(n15) edge node [above left] {} (n12)
(n12) edge node [above left] {} (n16.south)
(n12) edge node [above left] {} (n17.west)
(n8.south) edge node [above left] {} (n18.north)
(n16) edge node [above left] {} (n19)
;

\draw[->] (n6.east) -| (n8.north);
\draw (n10.east) -- node [fill=white] {qty}  (n15.west);

\end{tikzpicture}

\caption{High-level architecture of the integrated invoicing}
\label{fig:rims-architecture}
\end{figure}
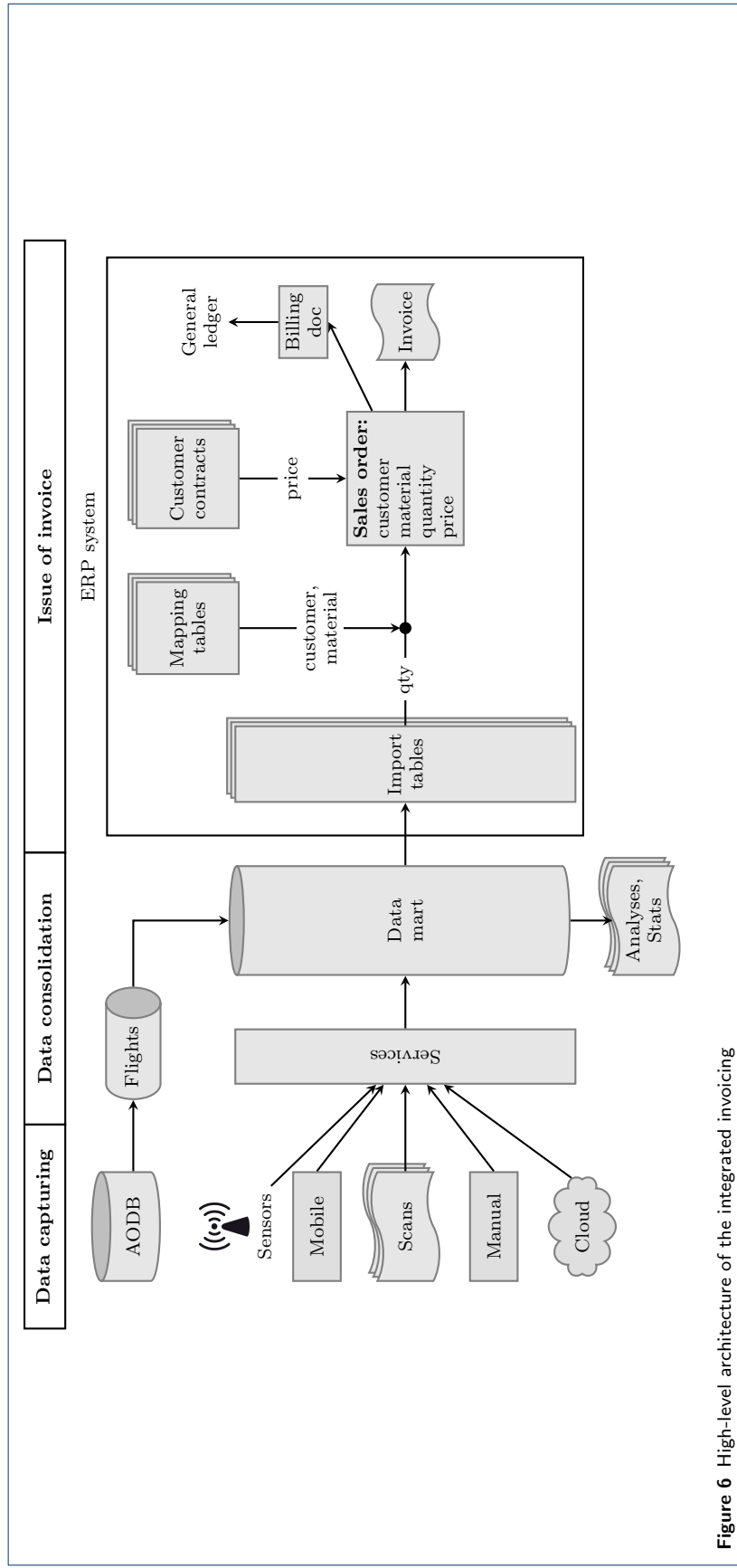
\end{landscape}

Figure \ref{fig:rims-architecture} shows the high-level architecture of integrated invoicing aligned to key steps of the invoicing process (i.e. data capturing, data consolidation, invoice issue). In this section, we describe each of these procedural steps and discuss their solutions in detail.

Digital transformation of the invoicing process may have its origin in automation of data capture. A ground handling agent usually provides services at different places: on the ramp (servicing area of the airport), inside the airport terminal, and in airport lounges. Clearly, the company may face challenges in collecting, consolidating and reporting all relevant data about performed services. Consequently, Swissport began automation of data capture. Related digitalization initiatives addressed particular service groups one by one, such as ramp, lounge and deicing services. After the majority of digitalization initiatives were implemented, the company developed their capacity to automatically collect and store information regarding customer services.
\begin{figure}[hbt]
\centering
\includegraphics[width=0.5\columnwidth]{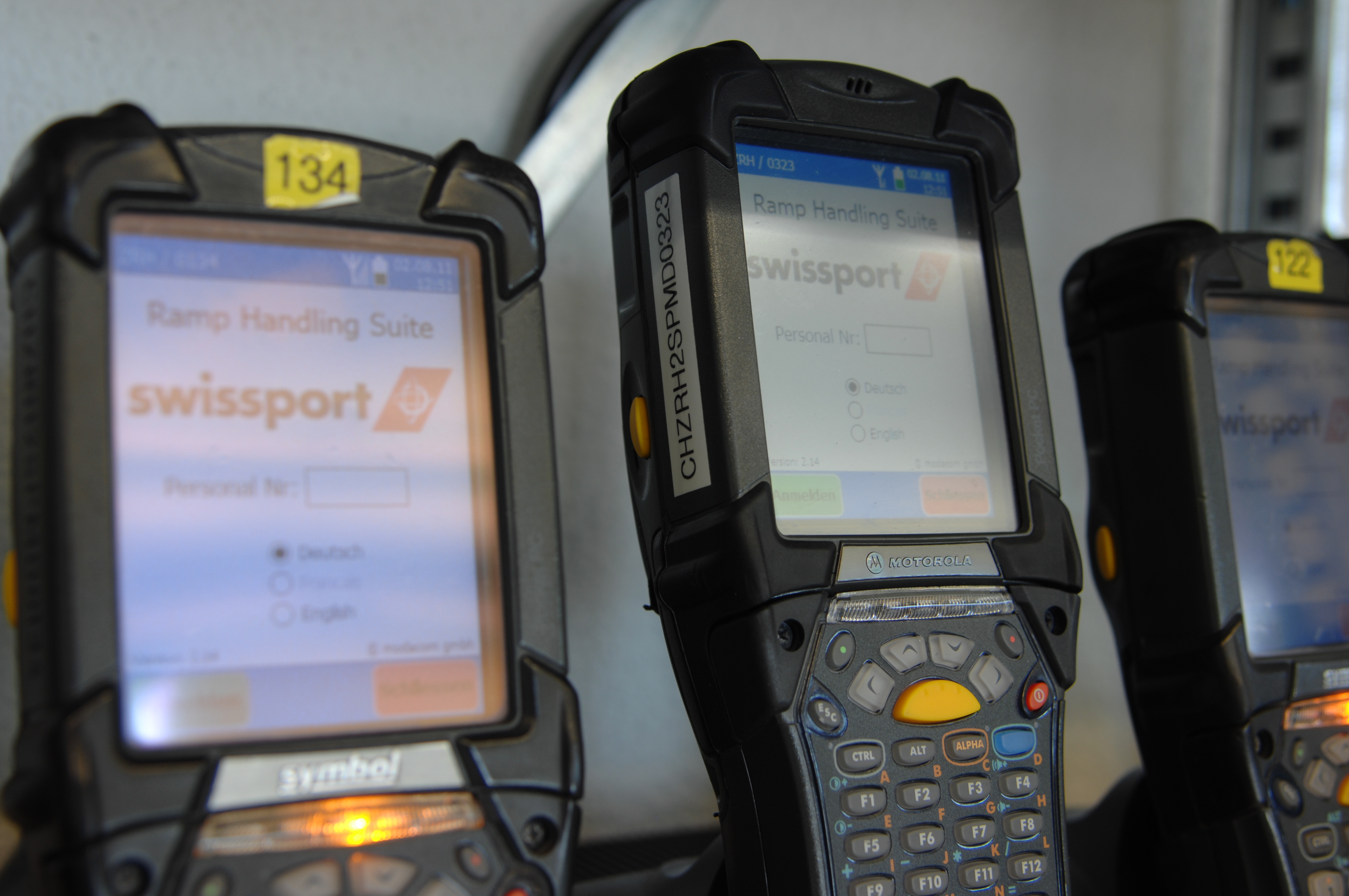}
\caption{Handhelds used to record ramp services (Photo courtesy of Swissport)}
\label{fig:handhelds}
\end{figure}

The following are some examples of how ground handling agents collect the flight and service data:
\begin{itemize}
\item flight data are imported from the airport operational database (AODB);
\item ramp services (e.g. loading equipment) are tracked by employees using hand-held devices (Fig. \ref{fig:handhelds});
\item ground support equipment (e.g. passenger boarding steps) employed during aircraft turnaround are tracked by means of GPS tracking and localization sensors;
\item the usage of deicing fluids and hot water is captured by sensors installed on de-icing trucks;
\item ground power units record the time required for aircraft batteries to recharge, using integrated sensors;
\item lounge entries are recorded using control gates when a passenger swipes the cards or scans the ticket;
\item lost and found services are recorded using mobile apps and dedicated desktop applications;
\item other infrequent or irregular services are captured on paper receipts (e.g. meal vouchers, taxi vouchers, handling work orders for private aviation flights, handling of VIP flights, special check-in counter reservations).
\end{itemize}
%
Once all the data on these services have been collected, they need to be consolidated and verified. In order to achieve this, the ground handling agent created a data-mart, which functions like a central repository for all flight and service data was created. The data-mart is designed to connect to multiple data sources at regular intervals, consolidate service data and, in the last step, link services to the corresponding flight. The service data can be then verified automatically using data-mining methods \cite{Ramkumar2013, Mikut2011}.

Finally, flight and service data are finally uploaded to the enterprise resource planning (ERP) system. The ERP system uses mapping tables to recognize the customer and the product or service (i.e. material) sold to the customer. Based on this information, the system automatically creates the sales order containing all individual items sold to the customer. All price data are included in the ERP system. The ERP system uses the price data to determine customer prices for the purchased items. Ultimately, the sales orders are billed and the invoices are transferred to the customer.

The ERP system integrates the invoicing as well as accounting capabilities, which is an advantage over other billing and invoicing solutions that are separate from the accounting systems. Once the invoice has been created and transferred to the customer, the system automatically creates corresponding positions on the customer account in the general ledger.

After all, this case study presents a best practice approach for implementing data intensive ERP systems. However, some contributions are novel and relate specifically to the field of airport ground operations:
\begin{itemize}
\item Key automated data flows across the ground handling agent and his external partners are introduced and discussed. In addition, key automated data flows inside the ground handling agent's organization are outlined and corresponding implications on the solution architecture are stated;
\item This case study explains why digital transformation projects in airport ground operations require high data quality and short data processing times. In addition, it shows how heterogeneous data collection procedures in airport ground operations have affected the overall design of the new invoicing solution;
\item This case study describes how specific circumstances in airport ground operations can be transferred into a standardized industry best practice solution without compromising on quality and performance. New procedures are lean, simple and, obviously, can be easily understood by anyone from outside the aviation industry;
\item The business scenario reported in this case study describes a digitalization initiative that has been successfully implemented without any interruptions in regular operations of the ground handling agent. Immediately after its introduction, the new invoicing solution has started to generate significant benefits towards costs, processing times and data quality, which is not obvious within a highly competitive environment of airport ground operations.
\end{itemize}

\section{Discussion}
\label{sec:Discussion}
The scenarios of digital transformation in airport ground operations and related technologies discussed in the previous section differ significantly in their maturity. So, we assessed these technologies and applications with respect to their impact and major improvements. In this section, we share key results of this assessment.

We rated all technologies and applications from the previous section according to their usefulness and maturity using a unified framework with clear evaluation criteria. On top of that, we developed hints and ideas for future research directions for rated technologies. We selected the technology readiness levels (TRLs) as a qualified measurement system to access maturity level of the related technologies.

TRLs were initially introduced by the National Aeronautics and Space Administration (NASA) to evaluate new technologies developed for space missions \cite{Mankins95} and, later on, generalized for all kinds of technology assessments\cite{Horizon2020}. TRLs are widely used in academia and industry, e.g. mandatory technology assessments for Horizon 2020\cite{Horizon2020} (a 80 billion Euro research program funded by the European Union), evaluation of cyber-physical systems \cite{Leitao16}, monitoring and evaluation of system development process \cite{Magnaye10}.

TRLs are based on a scale from 1 to 9. TRL 1 is the lowest level and TRL 9 is the highest. Then a technology is at TRL 1, there are preliminary results of scientific research are available and have been successfully translated into future research development. In the end, after a technology has been "mission proved", it achieves TRL 9 (cf. Table~\ref{tab:trl}).
\begin{table}[hbt]
\caption{Overview of technology readiness levels \cite{Mankins95}}
 \footnotesize
\begin{center}
\renewcommand{\arraystretch}{0.9}
\begin{tabularx}{\columnwidth}{lp{.75\columnwidth}}
\hline
TRL 1 &  Basic principles observed and reported \\ \hline
TRL 2 &  Technology concept and/or application formulated \\ \hline
TRL 3 &  Analytical and experimental critical function and characteristic proof of concept\\ \hline
TRL 4 &  Component or subsystem validated in laboratory environment \\ \hline
TRL 5 &  Component, subsystem or system validated in relevant  environment \\ \hline
TRL 6 &  Component, subsystem, system or prototype demonstrated in relevant end-to-end environment   \\ \hline
TRL 7 &  System prototype demonstrated in an operational environment \\ \hline
TRL 8 &  System completed and "mission qualified" through test and demonstrated in an operational environment \\ \hline
TRL 9 &  System "mission proven" through successful mission operations \\ \hline
\end{tabularx}
\end{center}
\label{tab:trl}
\end{table}

As previously mentioned, we have evaluated the technologies discussed in the previous section against the parameters for each TRL and then assigned a TRL rating based on reported applications, relevant research papers and our interviews with industry experts.
We summarized our key results in Tables~\ref{tab:trl_cp} (core processes) and~\ref{tab:trl_sp} (supporting processes). The applications reported in these tables are grouped by their position within the value chain of an airport ground handling agent (Figure~\ref{fig:valuechain}).
\begin{table}[hbt]
\caption{TRL ratings for technologies relating to core processes of a ground handling agent}
 \footnotesize
\begin{center}
\renewcommand{\arraystretch}{0.9}

\begin{tabularx}{\columnwidth}{p{.25\columnwidth}p{.5\columnwidth}p{.12\columnwidth}}
\hline
Business process 		& Technology or business scenario 			& TRL \newline rating \\ \hline
Passenger \newline handling 		& Indoor navigation						& 8 \\
					& Digital processing of irregularity vouchers	& 2 \\
					& Self-service check-in kiosks				& 9 \\
					& Self-boarding							& 9 \\
					& Smart wheelchairs						& 2 \\
					& Smart wearables						& 6 \\
					& Biometric services 						& 8 \\ \hline
Baggage \newline handling		& RFID baggage tags					& 8 \\
					& Automated baggage drop-off				& 9 \\
					& Digital bag tags 						& 8 \\
					& Self-tagging							& 8 \\
					& Lost luggage kiosks					& 9 \\
					& Real-time luggage tracking				& 2 \\
					& Advanced analytics in baggage handling	& 7 \\ \hline
Lounge services		& Lounge access gates					& 9 \\
					& New lounge ticket types					& 2 \\ \hline
\end{tabularx}
\end{center}
\label{tab:trl_cp}
\end{table}
\begin{table}[hbt]
\caption{TRL ratings for technologies relating to supporting processes of a ground handling agent}
 \footnotesize
\begin{center}
\renewcommand{\arraystretch}{0.9}

\begin{tabularx}{\columnwidth}{p{.25\columnwidth}p{.5\columnwidth}p{.12\columnwidth}}
\hline
Business process 		& Technology or business scenario 			& TRL \newline rating \\ \hline
Planning \& \newline Scheduling 	& Automated centralized planning 			& 9 \\
					& De-peaking 							& 9 \\
					& Shift trading 							& 7 \\ \hline
GSE \newline management 		& Automated GSE scheduling and routing 	& 6 \\\hline
HR \& \newline Training 			& Digital employee profiles 				& 6 \\
					& Web-based training 					& 9 \\
					& Integrated employee lifecycle management 	& 2 \\
					& Mobile apps and social-media for corporate communications & 9  \\ \hline
Financial process 		& Integrated invoicing 					& 9		 \\ \hline
Management process 	& Integrated KPI reporting					& 7 		 \\ \hline
IT process 			& Digital workplace and virtualization 		& 7		 \\ \hline
\end{tabularx}
\end{center}
\label{tab:trl_sp}
\end{table}

A considerable number of technologies has already achieved the highest TRL, e.g. self-service check-in kiosks, self-boarding, automated baggage drop-off. There is a wide range of scientific papers and industry reports available stating a broad commercial usage of these technologies (see the previous section for examples). Such technologies are "mission proven" and have been successfully operated in actual missions.

Another big set of technologies is currently at TRL 8, e.g. biometric services, digital bag tags, baggage self-tagging. These technologies have been tested and "mission qualified". We were able to find the evidence of the first commercial usage of these technologies, e.g. in terms of a successful pilot implementation.

We rated the technologies with system prototypes working in operational environment on full-scale realistic problems at TRL 7. Consider, for instance, the shift-trading. Some ground handling agents are using shift-trading systems with partial functionality. General engineering feasibility is fully demonstrated. Limited documentation is available. Furthermore, products covering partial functionality are available for commercial usage. In order to achieve the next TRL, system prototypes covering all key functionality need to be developed and integrated with existing operational software and hardware systems. These prototypes need to demonstrate full operational feasibility. For instance, several shift-trading systems are not fully integrated with time and attendance systems. The trades are put into the time and attendance systems manually by the back office staff or uploaded a-synchronically to the system as last step adjustments.

Consider now the usage of smart wearables in the airport ground operations. Smart wearables represent a bunch of technologies, e.g. smart watches and smart glasses. We rated these technologies at TRL 7 as well. We were able to find scientific papers and industry publications reporting well-functioning prototypes in operational environment, e.g. the pilot usage of smart glasses by ground handling staff at the lounge entry. Nonetheless, we were not able to find any evidence of a broad commercial usage of smart wearables in airport ground handling operations.

TRL 6 technologies have a full functioning prototype or a representational model, but they have not been tested in operational environment on full-scale realistic problems yet. Consider, for instance, the automated GSE scheduling and routing. Component and subsystem prototypes have been successfully demonstrated in operational environment; partially integrated with existing hardware and software systems. Products covering partial functionality for commercial usage are available, e.g. tracking of GSE vehicles and routing of de-icing trucks. In order to achieve the next TRL, end-to-end-software prototypes need to be developed and connected to existing systems conforming with target operating environment. New prototypes need to be tested in relevant environment and provide evidence of meeting expected performance. Scaling requirements need to be defined.

On top of that, we were able to identify a set of technologies at the very early stage of development. For instance, smart wheelchairs, real-time luggage tracking and digital processing of lounge vouchers are currently at TRL 2. Practical applications of these technologies have been successfully identified, but they are speculative. No experimental proof or detailed analysis are available to support the conjecture. Both analytical and laboratory studies are required to see if the technology is viable and ready to proceed through the further development process. For instance, a proof-of-concept model for digital processing of vouchers needs to be constructed in order to achieve TRL 3.

Overall, ground handling agents employ a broad set of high TRL technologies (TRL 6 and higher). Many of these technologies deliver cost reductions and efficiency improvements to  ground handling agents and are considered to be state-of-the-art technologies. Additionally, ground handling agents are seeking further differentiation and invest into new technologies (represented by low TRL technologies) in order to outpace competitors in terms of process efficiency and new revenue streams.

In addition to the necessary implementation in industry (including testing and improving new technologies, linking information flows etc.), some scientific challenges have to be solved. We have listed below some proposed research directions:
\begin{enumerate}

\item \textbf{Quantitative criteria for the evaluation of technologies}: Ground handling agents usually employ a number of quantitative criteria in order to evaluate technologies and make investment decisions relating to the use of new technologies.
For instance, low TRL technologies are basically evaluated against a number of technical criteria, e.g. minimal and maximal position error in indoor navigation. High TRL technologies are usually assessed in terms of economic prospects, e.g. expected return on investment, expected cost and staff reductions, new revenue streams based on new digital products or services. Unfortunately, we were not able to obtain reliable data on such technology assessments. Our interviews with industry experts indicate that ground handling agents are reluctant to publish such data because they consider them as part of the competitive advantage. We propose to develop generic evaluation criteria and procedures which can be used by ground handling agents to assess technologies and build reliable business cases.

\item \textbf{Incorporate uncertainty into forecasting and planning models}:
Several tasks and activities in airport ground handling are affected by uncertain and random events, e.g. flight delays, utilized capacity in flights, weather conditions and technical incidents. In contrast to another industries, the most forecasting and planning models used in the airport ground operations do not consider this kind of uncertainty explicitly (see forecasting of renewable generation and customer load in power systems \cite{GonzalezOrdiano18}, optionally using Big Data frameworks as Apache Spark \cite{GonzalezOrdiano18a}). By incorporating uncertainty into planning and forecasting models, we expect a significant increase of the robustness and reliability of developed plans.

\item \textbf{Multi-objective optimization}:
The majority of underlying optimization problems in planning and scheduling are formulated as optimization problems under technological or financial restrictions. The incorporated optimization function does not include such factors as customer satisfaction, employee satisfaction or ecological footprint. We believe that such optimization criteria can be incorporated into the existing optimization models at a low cost and without significant increase in recurring costs. On top of that, several optimization problems are formulated in terms of local optimization, e.g. reducing the number of check-in staff, and do not consider the optimization across the entire value chain. Hopefully, we were able to find some rare evidence of globally formulated optimization problems, e.g. considering the trade-off between minimum staff requirement against the quality of service.

\item \textbf{Use of collected data to generate new business models}:
The reported applications relating to the usage of enterprise data are focused basically on increasing process efficiency. We were able to find only a limited number of scenarios where ground handling agents were using available data to a greater extend or to generate new business models. For instance, globally acting ground handling agents have a large amount of data on flight delays and can use them to optimize flight schedule forecasts by exchanging these data across the local stations. On top of that, collected data may be a powerful resource for a new service offering to the airlines, e.g., passenger preferences for different airlines on the same routes and estimating the potential of promising direct routes to avoid changed flights.
Such tasks require prototypical projects of ground handling agents together with research institutions or consulting companies with a digitalization profile.
\end{enumerate}

\section{Conclusions}
\label{sec:Conclusions}
We found that the majority of digitalization initiatives in airport ground operations focus on operational improvements of airport ground operations. In the core processes of the airport ground operations value chain, the digitalization is mainly driven by cost pressures. The majority of customer-centric innovations contribute further cost reductions to ground handling agents, as well as benefits to the customers through time savings, improved service quality and transparency. Several passenger-related services have shifted towards digital self-service. Recent publications discuss several different digitalization scenarios in the passenger and baggage handling; however, some topics remain underrepresented, such as lounge service digitalization.

In supporting processes of the airport ground operations value chain, ground handling agents are trying to achieve further cost reductions by improving planning and scheduling procedures. We were able to find several papers on relevant methods and models, but very few discussed concrete solutions for implementation. Other supporting processes of the value chain, such as HR, IT, management and financial processes, were only infrequently discussed. Nonetheless, we identified sound digitalization scenarios relating to these processes, such as digital employee profiles, web-based trainings, mobile apps and social media for corporate communications, integrated KPI reporting, and integrated invoicing.

Ground handling agents rarely use new digital technologies to create new business models and disrupt markets; however, we identified various scenarios where digitalization has the potential to create new business models, such as with recent approaches to shift-trading, lounge ticketing and indoor navigation. Ground handling agents are engaging with business partners in order to create new revenue streams, such as baggage pickup and delivery at home, cooperation with taxis, duty-free shops in arrival lounges, lounge tickets sold by charter airlines and tour operators.

With these types of changes, the capacity to collect and analyze data has increased in importance. Ground handling agents are investing to consolidate, standardize and actively manage data pockets, in order to provide information on passenger flow, efficiency of baggage-handling systems, timeliness of flights, speed and effectiveness of passenger handling, and others. Ground handling agents aim to become more responsive and efficient by using this kind of information in operational and tactical decisions.

Based on our findings, we believe that the followings topics may be important future research directions in the area of airport ground operation digitalization: new cooperation modes across various business partners at airports, seamless travel, new digital business models, digital security solutions, multi-objective optimization, advanced forecasting and planning models with incorporated factors of uncertainty, general evaluation criteria for technology assessment in airport ground handling, and data collection and advanced analytics for operational decision making.


\begin{backmatter}

\section*{Declarations:}

\section*{Acknowledgments}
The authors also acknowledge careful English language editing by Rebecca Klady.

\section*{Competing interests}
The authors declare that they have no competing interests.

\section*{Funding}
This research did not receive any specific grant from funding agencies in the public, commercial, or not-for-profit sectors.


\bibliographystyle{bmc-mathphys} 
\bibliography{library}      

\end{backmatter}
\end{document}